\documentclass[12pt]{article}

\usepackage{times}
\usepackage{graphicx}
\usepackage{float}
\usepackage{subfigure}
\usepackage{array}
\usepackage{gensymb}

\usepackage{amsmath}
\usepackage[square,sort,comma,numbers]{natbib}
\usepackage{epstopdf} 

\usepackage{hyperref}

\usepackage{color,soul}
\usepackage[table,x11names]{xcolor}
\newcolumntype{M}[1]{>{\centering\arraybackslash}m{#1}}
\newcolumntype{P}[1]{>{\centering\arraybackslash}p{#1}}
\newenvironment{sciabstract}{%
\begin{quote}}
{\end{quote}}

\usepackage{float}

\newcounter{lastnote}

\title{Joule-Thomson expansion in AdS black hole with a global
monopole}

\author
{Ahmed Rizwan C.L.$^\ast$, Naveena Kumara A., Deepak Vaid. and Ajith K.M.\\
\\
\scriptsize{ Department of Physics, National Institute of Technology Karnataka (NITK)}\\
\scriptsize{Surathkal, Mangaluru - 575025, India}\\
\normalsize{$^\ast$E-mail:  ahmedrizwancl@gmail.com}
}

\date{}

\begin{document}

\maketitle

\begin{sciabstract}
{\bf{Abstract}} 

In this paper, we investigate the Joule Thomson effects for AdS black holes with a global monopole. We study the effect of the global monopole parameter $\eta$ on the  inversion temperature and isenthalpic curves.  The obtained result is compared with Joule Thomson expansion of van der Waals fluid and the equivalence were noted. Phase transition occuring in the extended phase space of this black hole is analogous to van der Waals gas. Our study shows that global monopole parameter $\eta$ plays a very important role in Joule Thomson expansion.

\end{sciabstract}

\tableofcontents

\section{Introduction}
\label{intro}
Black hole thermodynamics has become an important area of research in recent years due to its immense impact on understanding quantum gravity \citep{Wald2001}, \citep{Page2005}, \citep{Hawking1983}, \citep{Bardeen1973}, \citep{Bekenstein1972}, \citep{Bekenstein1973}, \citep{Bekenstein1974}. This importance lies in the fact that the microscopic structure of the black hole is not well understood in the framework of quantum gravity. In this regard the study of black hole becomes inseparable topic since its theoretical domain includes quantum mechanics and general relativity. As a ray of hope in this cluttered scenario, classical thermodynamic study of asymptotically anti-de Sitter black hole emerged as the successor, equipped with critical phenomena and phase structure  results backing the AdS/CFT correspondence \citep{Maldacena1999}. This duality enabled us to treat the topics like quark gluon plasma and various systems in condensed matter physics with more confidence and new insights.   Among all,  a pathbreaking correspondence has been established by \citep{Davies1978},\citep{Chamblin1999}, \citep{Chamblin:1999hg}  between the phase diagrams of charged AdS black hole and van der Waals fluids \citep{Kubiznak2012}. 

   In the early stage of the developement of black hole thermodynamics the mass of a black hole has been identified with its energy. In another approach, the cosmological constant is treated as thermodynamic pressure  with the aid of extended phase space and thus interpreting the mass of the black hole as  chemical enthalpy rather than its energy  \citep{Kubiznak2017}\citep{Dolan2011}. The study of black hole thermodynamics in the extended phase space enables us to identify new thermodynamic variable, volume of the black hole which is conjugate to pressure. These new identifications lead to the emergence of thermodynamic phase transitions with the condensed matter counterpart like van der Waals fluid which is of our interest in this paper. Thermodynamics of various AdS black holes showing similar thermodynamic behaviour lead to the conclusion that this phenomenon is a universal feature \citep{Gunasekaran:2012dq}. Again classical thermodynamic concept of a heat engine is applied to black hole thermodynamics and a way to convert mechanical energy from the black hole is found \citep{0264-9381-31-20-205002}.

   An interesting extension of black hole thermodynamics into the regime of Joule-Thomson  (JT) expansion is carried out recently in the work of {\"{O}}kc{\"{u}} and Aydıner \cite{Okcu2017}. They have studied JT expansion of charged AdS black hole and later in Kerr AdS black hole \citep{Okcu2018}, wherein inversion curve and isenthalpic curves are studied and compared with the van der Waals fluid.  As an extension this method is applied to a spacetime in $d$ dimension \citep{Mo2018}. In another application the corrections of the quintessence on the Joule-Thomson (JT) effect of the Reissner Nordstrom anti de Sitter (RNAdS) black hole is reported \citep{Ghaffarnejad2018}. Recently, effects of  Lovelock gravity and JT expansion in charged Gauss-Bonnet black hole has been studied \citep{Mo:2018qkt} \citep{Lan:2018nnp}.

    It is a well known fact that the phase transitions in the evolution of the early Universe can give rise to topological defects like global monopole \citep{Kibble1976} \citep{Vilenkin1985}. Historically the concept of magnetic  monopole was first introduced by Dirac in 1931. Monopoles are formed from breaking of a gauge symmetry  which carries a unit of magnetic flux \citep{Vilenkin1994}. The formation of monopoles can be compared with the elementary particle formation where  SO(3)  gauge symmetery is broken. In a seminal work, Barriola and Vilenkin \citep{Barriola1989} obtained the black hole metric with a global monopole which has a distinct topological structure compared to Schwarzschild black hole. The gravitational fields associated with global monopole lead to anisotropies in the microwave background radiation and this fluctuations created  later evolve into galaxies and clusters.
 The Significance of the global monopole paramater in holographic superconductivity  studied  by Chen et al., showed that superconducting transition  depends on the monopole parameter \citep{Chen:2009vz}. Recently the thermodynamics of this blackhole is studied by Deng et al., \citep{Deng2018} , where it is reported that the presence of global monopole plays a vital role in  van der Waals like phase transition. 

This paper is organised as follows. In section \ref{monopole} we discuss the details of the charged AdS black hole metric and its thermodynamic properties. This is followed by a section on (sect. \ref{secJT}) Joule Thomson expansion for van der Waals fluids where we briefly review the well known results in classical thermodynamics. In section \ref{JTMonopole}  we explore JT expansion for charged AdS blackhole with global monopole. We derive expression for inversion temperature and study the nature of inversion curve and isenthalpic plots. We conclude that section by calculating the ratio between inversion temperature $T_i$ and critical temperature $T_c$. The ratio is compared to the earlier calculations for Kerr-AdS black hole \citep{Okcu2017} and charged AdS black hole\citep{Okcu2018}. In the last section we present our results.    
\section{The charged AdS black hole with global monopole}\label{monopole}
\subsection{The Metric}
In this section, we briefly review the thermodynamic properties of charged AdS blackhole with global monopole. The simplest model that gives rise to the global monopole is described
by the Lagrangian \citep{Barriola1989},
\begin{equation}
\mathcal{L}_{gm}=\frac{1}{2}\partial _\mu \Phi ^a \partial  ^\mu {\Phi ^*} ^a-\frac{\gamma}{4}\left( \Phi ^a{\Phi ^*}^a-\eta _0 ^2\right)^2,
\end{equation}
where $\Phi ^a$ is a multiplet of scalar field, $\gamma$ is a constant and $\eta$ is the energy scale of symmetry breaking.  In four dimensional space AdS blackhole  with global monopole is defined with the metric
\begin{equation}
d\tilde{s}^2=-\tilde{f}(\tilde{r})d\tilde{t}^2+\tilde{f}(\tilde{r})^{-1}d\tilde{r}^2+\tilde{r}^2 d\Omega ^2
\end{equation}
where $d\Omega ^2=d\theta ^2 +\sin ^2 \theta d\phi ^2$ and $\tilde{f}(\tilde{r})$ is given by,
\begin{equation}
\tilde{f}(\tilde{r})=1-8\pi \eta _0^2 -\frac{2\tilde{m}}{\tilde{r}}+\frac{\tilde{q}^2}{\tilde{r}^2}+\frac{\tilde{r}^2}{l^2}~~,~~\tilde{A}=\frac{\tilde{q}}{\tilde{r}}d\tilde{t}.
\end{equation}
In the above equations $\tilde{m}$, $\tilde{q}$ and $l$ are the mass parameter, electric charge parameter and AdS radius of the black hole, respectively. We make the following coordinate transformation,

\begin{equation}
\tilde{t}=(1-8\pi \eta _0^2)^{-1/2}t~~,~~\tilde{r}=(1-8\pi \eta _0^2)^{1/2}r
\end{equation}
and introduce new parameters
\begin{equation}
m=(1-8\pi \eta _0^2)^{-3/2}\tilde{m}~~,~~q=(1-8\pi \eta _0^2)^{-1}\tilde{q}~~,~~\eta ^2=8\pi \eta _0^2.
\end{equation}
Finally we have the line element
\begin{equation}
ds^2=-f(r)dt^2+f(r)^{-1}dr^2+(1-\eta ^2)r^2 d\Omega ^2
\end{equation}
\begin{equation}
f(r)=1 -\frac{2m}{r}+\frac{q^2}{r^2}+\frac{r^2}{l^2}~~,~~A=\frac{q}{r}dt.
\end{equation}
In terms of the corresponding parameters the electric charge and ADM mass are given by
\begin{equation}
Q=(1-\eta ^2)q~~,~~M=(1-\eta ^2)m.
\end{equation} 
\subsection{The Thermodynamics}
The largest root of $f(r_+)=0$ gives the black hole event horizon, where $r_+$ is the location of the event horizon of the black hole. One can use this to express the ADM mass as follows,
\begin{equation}\label{eq9}
M=\frac{(1-\eta ^2)}{2}r_++\frac{Q^2}{2r_+(1-\eta ^2)}+\frac{r_+^3(1-\eta ^2)}{2l^2}.
\end{equation}
Note that the first law of thermodynamics and Smarr relation holds good for this blackhole, which reads as follows
\begin{equation}
dM=TdS+\Phi dQ+VdP~~,~~M=2(TS-PV)+\Phi Q.
\end{equation} 
The entropy $S$ of the blackhole is obtained from the area $A_{bh}$ of the event horizon,
\begin{equation}\label{eq11}
S=\frac{A_{bh}}{4}=\pi (1-\eta ^2)r_+^2.
\end{equation}
In the extended phase space the cosmological constant corresponds to the thermodynamic variable pressure and its conjugate quantity corresponds to the thermodynamic volume,
\begin{equation}
P=-\frac{\Lambda}{8\pi}=\frac{3}{8\pi l^2}~~,~~V=\frac{4}{3}\pi (1-\eta ^2)r_+^3.
\label{eqP}
\end{equation}
The Hawking temperature is obtained as follows
\begin{equation}
T=\left( \frac{\partial M}{\partial S}\right) _{P,Q}=\frac{1}{4\pi r_+}\left(1+\frac{3r_+^2}{l^2}-\frac{Q^2}{(1-\eta ^2)^2r_+^2}\right).
\label{eqT}
\end{equation}
It is clear from equations (\ref{eq9}), (\ref{eq11}), (\ref{eqP}) and (\ref{eqT}) that the thermodynamic variables of the black hole under investigation depends on $\eta$. Hence it is reasonable to suspect that the thermodynamic behavior of the black hole will rely on the symmetry breaking scale. 
\begin{figure*}[ht!]
	\subfigure[]
	{
	\includegraphics[width=0.5\textwidth]{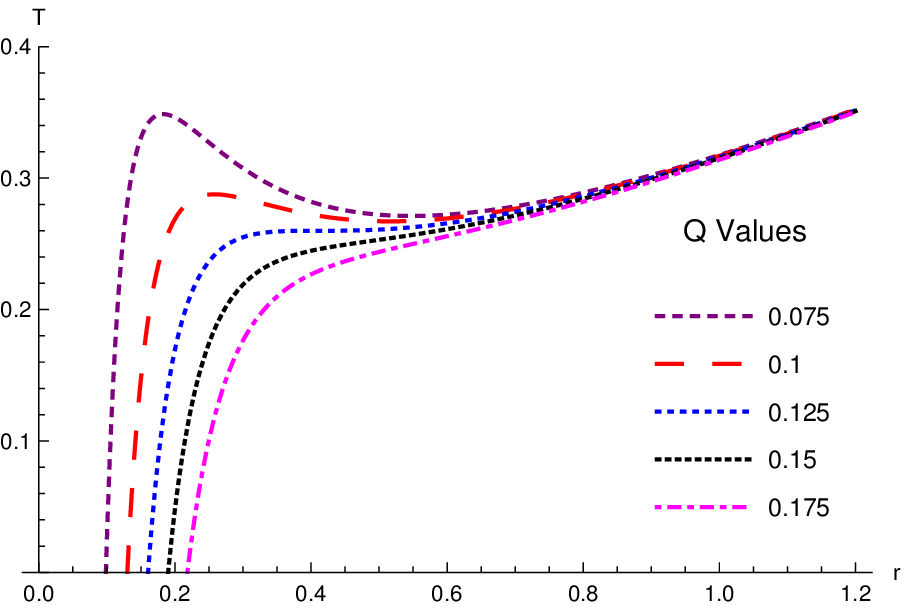}
		\label{Tr}}
	\subfigure[]
	{
		\includegraphics[width=0.5\textwidth]{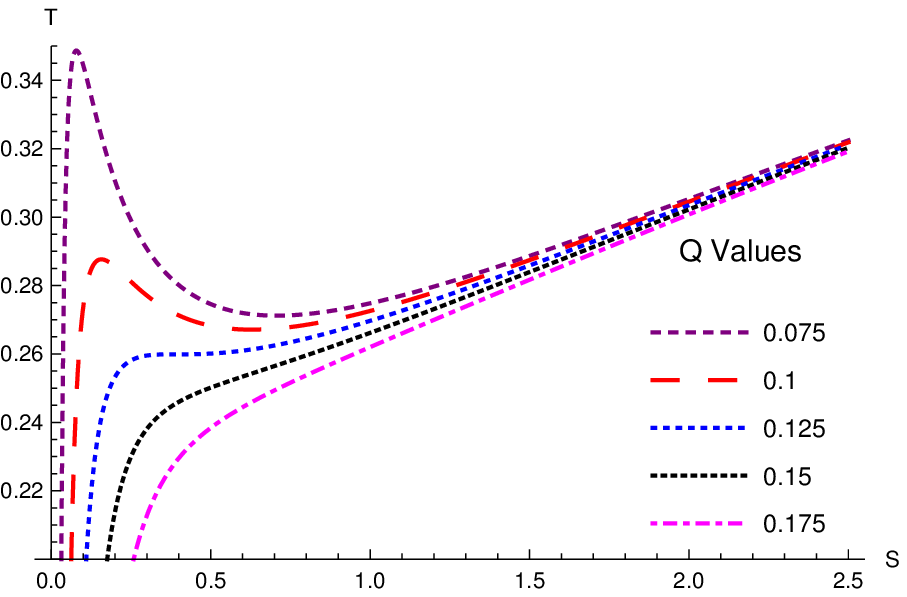}
		\label{TS}
	}
\\
  \caption{Plots of $T$ versus $r_+$ and  $T$ versus $S$ for different values of $Q$ and for $\eta =0.5$. These plots shows the behavior of Hawking temperature.}
   \end{figure*}
As the slope of the $T-S$ graph is related to specific heat, it's positive and negative values are related to the stability and unstability of the system with respect to fluctuations. The plots  \ref{Tr} and \ref{TS} shows that there exists a critical point which indicates the phase transition. At this stage we recall from the literature that in a canonical ensemble where charge is fixed, the asymtotically AdS blackholes shows a first order phase transition similar to van der Waals fluids terminating in a second order critical point.   
\subsection{Equation of state}
Combining the expressions for Hawking temperature (eqn. \ref{eqT}) and the thermodynamic pressure (eqn. \ref{eqP}), we get the {\emph{equation of state} for a charged AdS blackhole with global monopole as, 
\begin{equation}
P=\frac{T}{2r_+}-\frac{1}{8\pi r_+^2}+\frac{Q^2}{8\pi(1-\eta ^2)^2r_+^4}.
\label{eqstate}
\end{equation}
The above geometric equation of state is converted to physical equation of state under the dimensional analysis ground using the following scalings,
\begin{equation}
\tilde{P}= \frac{\hbar c}{l_P^2}P~~,~~\tilde{T}=\frac{\hbar c}{k}T,
\end{equation}  
where $l_P$ is the Planck length. Comparing geometric equation of state with van der Waals equation, one can relate the specific volume $v$ with the horizon radius $r_+$ as $v=2l_P^2r_+$. Using these relations the physical equation of state is obtained to be
\begin{equation}
P=\frac{T}{v}-\frac{1}{2\pi v^2}+\frac{2Q^2}{\pi(1-\eta ^2)^2v^4}.
\label{P monopole}
\end{equation}
At the critical point, 
\begin{equation}
\left(\frac{\partial P}{\partial v}\right)_T=\left(\frac{\partial ^2 P}{\partial v^2}\right)_T=0.
\end{equation}
Using this, the critical quantities are obtained as
\begin{equation}
P_c=\frac{(1-\eta ^2)^2}{96 \pi Q^2}~~,~~v_c=\frac{2\sqrt{6}Q}{1-\eta ^2}~~,~~T_c=\frac{1-\eta ^2}{3\sqrt{6}\pi Q^2}.
\end{equation}
From these formulas, it is clear that the presence of global monopole term affects the critical pressure $P_c$, critical temperature $T_c$ and the critical volume $v_c$ compared to the Reissner-Nordstr{\"{o}}m AdS black hole; $P_c$ and $T_c$ decreases while $v_c$ increases.
\section{Joule Thomson expansion}\label{secJT}

\subsection{Joule-Thomson effect}
Joule-Thomson effect is an irreversible adiabatic expansion of a gas when the gas is pushed through a porous plug. In this process,  a non-ideal gas undergoes continuous throttling process which leads to a temperature change in the final state. When the gas in the higher pressure side having pressure $P_i$ and temperature $T_i$ is made to expand through a porous plug, the gas passes through dissipative non-equilibrium states due to the friction between the gas and the plug. Usual thermodynamic coordinates cannot be used to define  these non-equilibrium states, but it is found that enthalpy which is the sum of internal energy and product of pressure volume remains same in the final state \citep{Zemansky2011}. So a state function called \emph{enthalpy} $ H=U+PV$ is defined, which remains unchanged in the end states,
\begin{equation}
H_i= H_f
\label{H}
\end{equation}
It is not entitled to say that enthalpy is a constant during this process, since  enthalpy is not defined when gas traverses  non-equilibrium states. The set of discrete points in the phase diagram initial point $(P_i, T_i)$ and all other points $P_f$s and $T_f$s representing equilibrium states of some gas having same molar enthalpy $(h)$ at initial and all the final equilibrium states. These discrete points corresponding to same molar enthalpy lies on a smooth curve known as \emph{isenthalpic curve}. To summarise an isenthalpic curve is the locus of all points with the same molar enthalpy representing intial and final equilibrium states. A set of such curves can be obtained for different values of enthalpy.

The slope of an isenthlpic curve on $T-P$ plane is called the \emph{Joule Thomson coefficient} $\mu _J$.
\begin{equation}
\mu _J=\left( \frac{\partial T}{\partial P}\right) _H.
\end{equation}
Joule Thomson coefficient is zero at the maxima of the isenthalpic curve. The locus of such points is called inversion curve. The interior of the inversion curve where the gradient of isenthalps ($\mu _J$) positive is called region of cooling and the exterior where $\mu _J$ is negative  called the region of heating. 
The differential of molar enthalpy is given by
\begin{equation}
dh=Tds+vdP.
\label{dh}
\end{equation}
We recall the second $TdS$ equation in classical thermodynamics \citep{Zemansky2011}
\begin{equation}
TdS=C_PdT-T\left( \frac{\partial v}{\partial T}\right) _PdP.
\label{tds}
\end{equation}
Using equation(\ref{tds})  in equation(\ref{dh}),
\begin{equation}
dT=\frac{1}{C_P}\left[ T\left( \frac{\partial v}{\partial T}\right) _P-v\right]dP+\frac{1}{C_P}dh.
\end{equation}
Which gives
\begin{equation}
\mu _J=\left( \frac{\partial T}{\partial P}\right) _H=\frac{1}{C_P}\left[ T\left( \frac{\partial v}{\partial T}\right) _P-v\right].
\end{equation}
As $\mu _J=0$ defines the inversion temperature we have
\begin{equation}
T_i=V\left( \frac{\partial T}{\partial v}\right)_P.
\label{inT}
\end{equation}

\subsection{van der Waals fluid}
van der Waals gas is the simplest model used to explain the behaviour of the real gases, which departs from the ideal gas discription with richer outcomes as it includes the intermolecular interaction and the non zero size of the molecules. The equation of state for a van der Waals gas is given by,
\begin{equation}
\left( P+\frac{a}{V_m^2}\right) \left( V_m-b\right)=RT.
\label{van}
\end{equation}
Here the constants $a$ and $b$ parameterizes the strength of the intermolecular interaction and the volume excluded due to the finite size of the molecule, respectively. The equation of state reduces to ideal gas equation under the limit $a$ and $b$ both set to zero. We used $V_m$ for molar volume, which is simply $V$ for one mole of substance.
\begin{figure}[H]
	\subfigure[]
	{
	\includegraphics[width=0.5\textwidth]{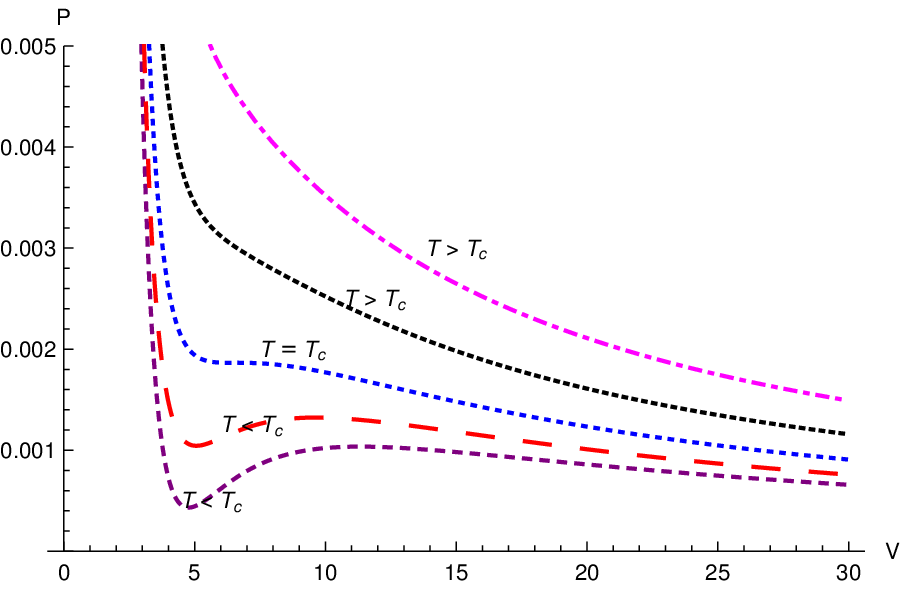}
		\label{PV van}}
	\subfigure[]
	{
		\includegraphics[width=0.5\textwidth]{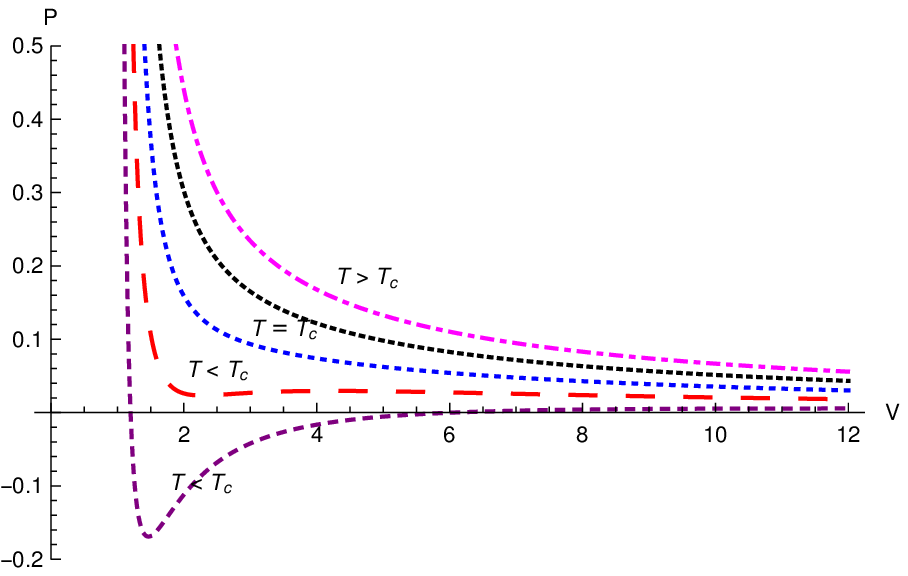}
		\label{PV monopole}
	}
\caption{Fig \ref{PV van} is isotherms of the van der Waals gas with temperature decreases from top to bottom. Fig \ref{PV monopole} shows $P-v$ diagram of charged AdS blackhole with global monopole where the parameters are choosen to be $Q=1$ and $\eta =0.5$. Similar $P-v$ diagrams can be obtained for different values of $\eta$. Variation of $\eta$ does not changes the nature of the $P-v$ diagram even though it changes the critical parameters.}
   
\end{figure}
To calculate the critical points namely the temperture $T_c$, pressure $P_c$ and volume $V_c$ we rearrange the equation of state  (equation \ref{van}) for $P$ as follows,
\begin{equation}
P=\frac{RT}{V -b}-\frac{a}{V^2}.
\end{equation} 
Using this equation the $P-v$ isotherms are plotted for van der Waals gas in fig.  \ref{PV van}. Fig \ref{PV monopole} is the $P-v$ diagram of the charged AdS black hole, which is obtained from equation (\ref{P monopole}), is  having a typical behavior of a van der Waals fluid. In both these graphs below a certain point called critical point there are inflection points and above that a monotonic behavior is displayed. This is a general result for AdS black holes \citep{Kubiznak2012}.
At the critical point $\left( \frac{\partial P}{\partial V}\right) _T=\left( \frac{\partial ^2 P}{\partial V^2}\right) _T=0 $, which gives
\begin{equation}
V_c=3b\ \ , \ \ T_c=\frac{8a}{27Rb}\  \,\ \ P_c=\frac{a}{27b^2}.
\end{equation}
The internal energy of van der Waals gas is given by \citep{Landau1980}
\begin{equation}
U(T,v)=\frac{3}{2}k_BT-\frac{a}{v}.
\end{equation}
Making a Legendre transformation $H=U+PV$ we obtain the expression for enthalpy,
\begin{equation}
H(T,v) =\frac{3}{2}k_BT+\frac{k_BTv}{v-b}-\frac{2a}{v}.
\end{equation}
The inversion temperature is calculated from equation (\ref{inT})
\begin{equation}
T_i=\frac{1}{k_B}\left( P_iv-\frac{a}{v}+\frac{2ab}{v^2}\right)
\label{t1}
\end{equation} 
and from equation of state (equation \ref{van}) we have
\begin{equation}
T_i=\frac{1}{k_B}\left( P_iv-P_ib+\frac{a}{v}-\frac{ab}{v^2}\right).
\label{t2}
\end{equation}
\begin{figure}[H]
	\subfigure[]
	{
\includegraphics[width=.5\textwidth]{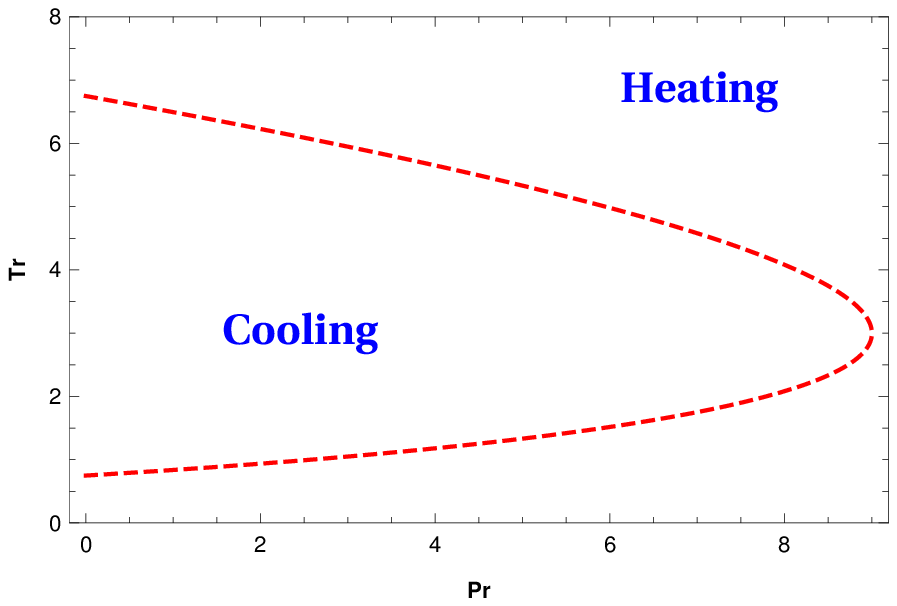}
		\label{inv}
	}
    \subfigure[]
    {
\includegraphics[width=.5\textwidth]{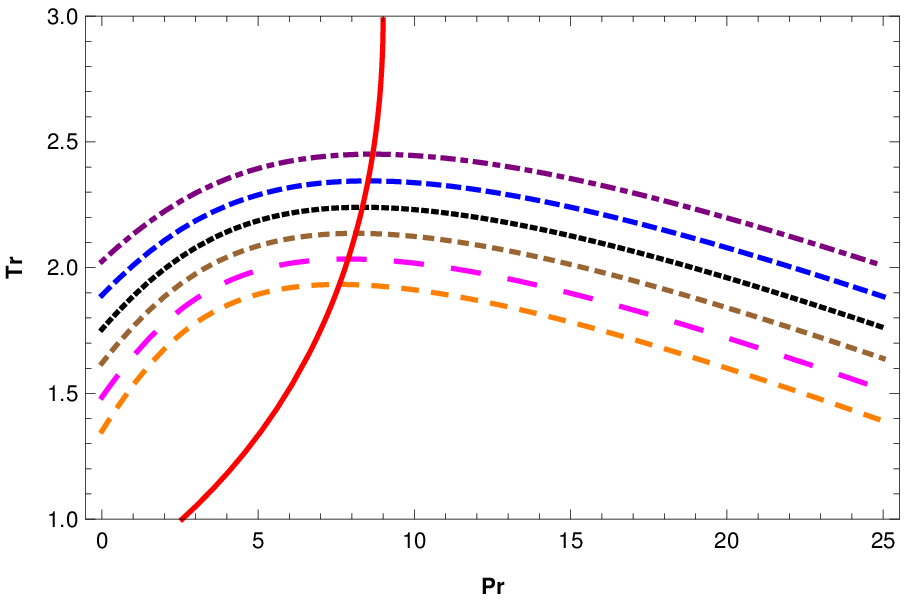}
        \label{together}
     }
  \caption{In fig. \ref{inv} inversion curve separating the regions of heating and cooling are shown. In fig \ref{together} isenthalpic curves for different values of enthaly is plotted along with the lower half of inversion curve for the van der Waals gas. While ploting this, we worked with dimensionless coordinates i.e., reduced pressure $P_r=P/P_c$ and reduced temperature $T_r=T/T_c$.}
\end{figure}
Equations (\ref{t1}) and (\ref{t2}) gives
\begin{equation}
Pbv^2-2av+3ab=0.
\end{equation}
Solving the above equation for $v$ and substituting in equation of state (equation \ref{van})
we obtain
\begin{equation}
T_i=\frac{2 \left(5a-3b^2P_i\pm 4\sqrt{a^2-3ab^2P_i}\right) }{9bk_B}.
\end{equation}
Using this we plot the inversion curves fig. \ref{inv}. In fig. \ref{together} isenthalpic and inversion curves are shown together. For the sake of comparison later with the isenthalpic-inversion curve of black hole we have taken only lower half of $T_i$.
\section{Joule Thomson expansion of charged AdS blackhole with monopole term}\label{JTMonopole}
In this section we study the JT expansion of charged AdS black hole with monopole term. Because of the treatment of black hole mass equivalent to enthalpy in extended phase space the isenthalpic plots are replaced by constant mass plots in this case (for brevity we still call it as isenthalpic curve).

Recall the expression for Joule Thomson coefficient 
\begin{equation}
\mu _J =\left( \frac{\partial T}{\partial P}\right) _M=\frac{1}{C_P}\left[ T\left( \frac{\partial V}{\partial T}\right) _P-V\right].
\end{equation}
From this we obtain the the invesion temperature
\begin{equation}
T_i=V\left( \frac{\partial T}{\partial V}\right) _P.
\label{eqJT}
\end{equation}
For this we rewrite the equation of state interms of $V$ as follows
\begin{equation}
\begin{split}
T=&\left( \frac{(1-\eta ^2)}{48\pi ^2}\right)^{1/3} \frac{1}{V^{1/3}}+P \left( \frac{6}{\pi (1-\eta ^2)}\right) ^{1/3}V^{1/3}\\
& -\frac{Q^2}{3(1-\eta ^2)}\frac{1}{V}.
\end{split}
\end{equation}
Substituting this into equation (\ref{eqJT}) we have the inversion temperature 
\begin{align}
T_i=&-\frac{1}{6} \left( \frac{(1-\eta ^2)}{6\pi ^2}\right)^{1/3} \frac{1}{V^{1/3}}+P \left( \frac{2}{9\pi (1-\eta ^2)}\right) ^{1/3}V^{1/3}\\
&+\frac{Q^2}{3(1-\eta ^2)}\frac{1}{V} \nonumber \\
=&\frac{Q^2}{4\pi r_+^3(1-\eta ^2)^2}+\frac{2}{3}Pr_+-\frac{1}{12\pi r_+}.
\label{eqstate1}
\end{align}
From equation(\ref{eqstate}) we have
\begin{equation}
T_i=-\frac{Q^2}{4\pi r_+^3(1-\eta ^2)^2}+2Pr_++\frac{1}{4\pi r_+}.
\label{eqstate2}
\end{equation}
From equation (\ref{eqstate1}) and equation (\ref{eqstate2}) we get
\begin{equation}
8\pi (1-\eta ^2)^2Pr_+^4+2(1-\eta ^2)^2r_+^2-3Q^2=0.
\label{root eq}
\end{equation}
\begin{figure}[!]
\begin{center}
\includegraphics[scale=1]{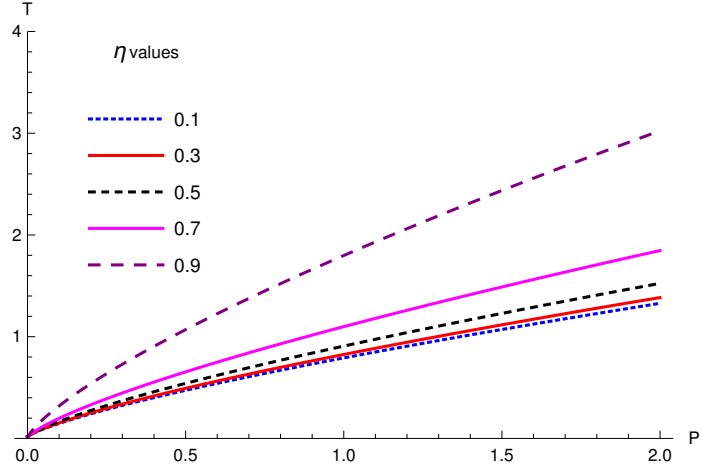}
\caption{Effect of monopole term $\eta$ on inversion curve. Here we have chosen  different values of monopole  ($\eta=0$ to $0.9$ in steps) by  keeping charge $Q$ fixed.}
\label{linear eta}
\end{center}
\end{figure}
Solving the above for $r_+$ and choosing the following appropriate root, 
\begin{equation}
r_+=\frac{1}{2 \sqrt{2 \pi }} \sqrt{\frac{\sqrt{ \left(\left(1-\eta ^2\right)^2+24 \pi  P Q^2\right)}}{\left(1-\eta ^2\right) P}-\frac{1}{P}}.
\end{equation}
Substituting this root into equation (\ref{eqstate2})
\begin{equation}
T_i=\frac{\sqrt{P_i} \left( 1+\frac{16\pi P_iQ^2}{(1-\eta ^2)^2 }-\frac{\sqrt{24P_i\pi Q^2+(1-\eta ^2)^2}}{(1-\eta ^2)}\right)}{\sqrt{2\pi} \left( -1+\frac{\sqrt{24P_i\pi Q^2+(1-\eta ^2)^2}}{(1-\eta ^2)} \right)^{3/2}}.
\end{equation}
From this equation the inversion curves are plotted for different values of $\eta$ (fig \ref{linear plots}). From the graphs one can infer that the JT coefficient $\mu _J$ is sensitive to $\eta$ values i.e., $\mu _J$ increases with $\eta$. In fig. (\ref{linear eta}), this inference is depicted taking different $\eta$ values in the same plot for a fixed $Q$ value.

By demanding $P_i=0$, we obtain $T_i^{min}$
\begin{equation}
T_i^{min}=\frac{(1-\eta ^2)}{6\sqrt{6}\pi Q}.
\end{equation}
Using this we calculate the ratio between $T_i^{min}$ and $T_c$ as
\begin{equation}
\frac{T_i^{min}}{T_c}=\frac{1}{2}.
\end{equation}
This is an interesting result which matches with the earlier established resultS for the charged AdS blackhole \citep{Okcu2017}, \cite{Okcu2018}. At the end of this study, we plot isenthalpic curves for various combinations of $\eta$ and $Q$ in the $T-P$ plane. Inverse points $(T_i, P_i)$ on $T-P$ plane seperates heating phase from the cooling phase of JT expansion. Recall that  isenthalpic curve in this case is not a plot with constant enthalpy, rather constant mass. The crossing diagram between inversion and isenthalpic curve shown in fig. ($\ref{crossing}$) displays the sensitivity of inverse points $(T_i,P_i)$ for the different values of $\eta$ and $Q$. All our calculations and graphs shows that when global monopole parameter is zero, the results nicely reduces to the earlier studies on JT expansion of charged AdS black holes.

\begin{figure*}[!]
	\subfigure[]
	{
		\includegraphics[width=0.5\textwidth]{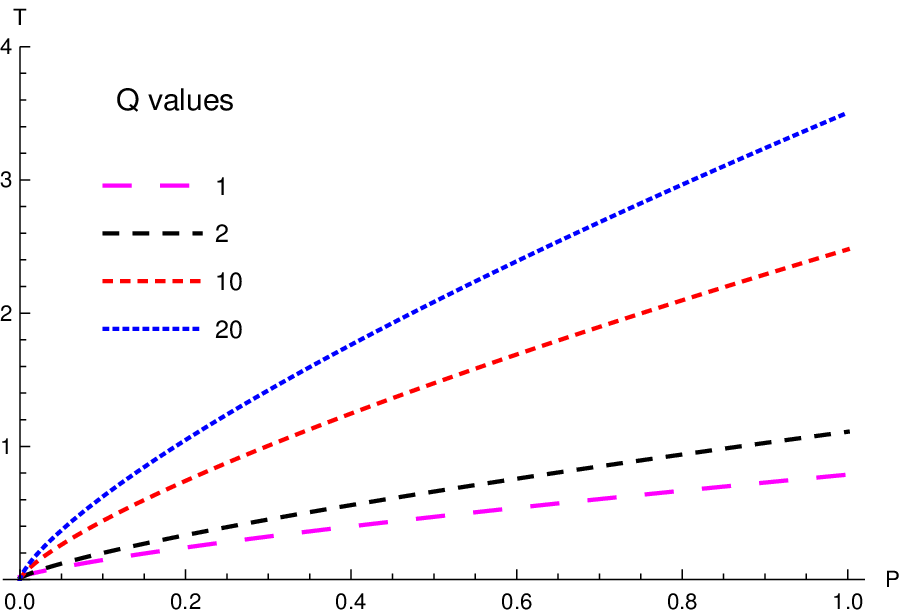}
		\label{fig:1}
	}
    \subfigure[]
    {
        \includegraphics[width=0.5\textwidth]{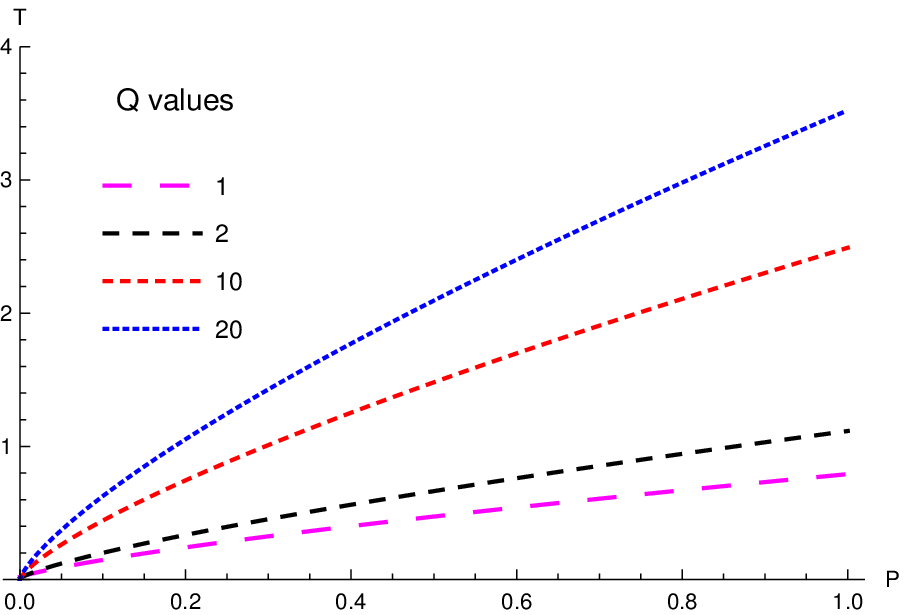}
        \label{fig:2}
     }
     \\
    \subfigure[]
    {
    	\includegraphics[width=0.5\textwidth]{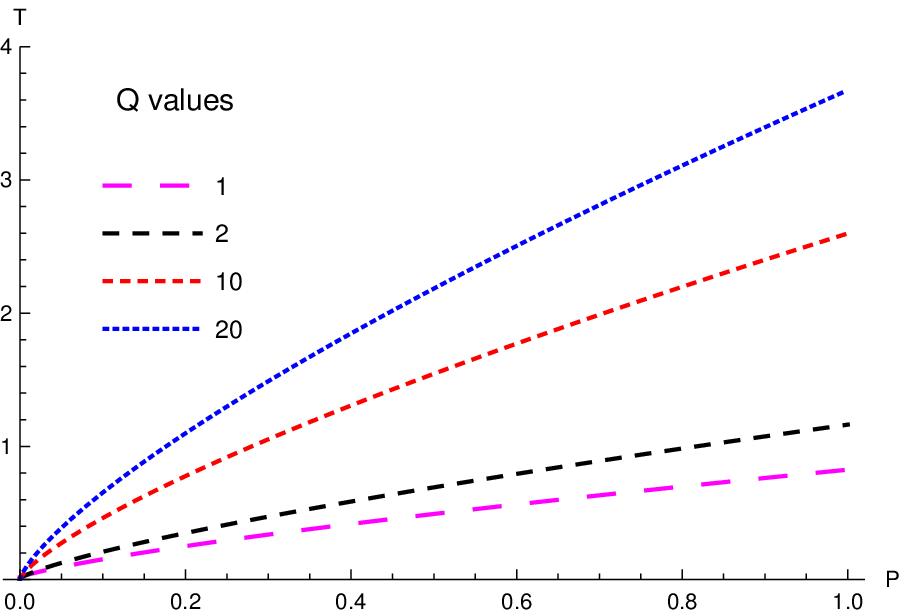}
    	\label{fig:3}
    }
    \subfigure[]
    {
    	\includegraphics[width=0.5\textwidth]{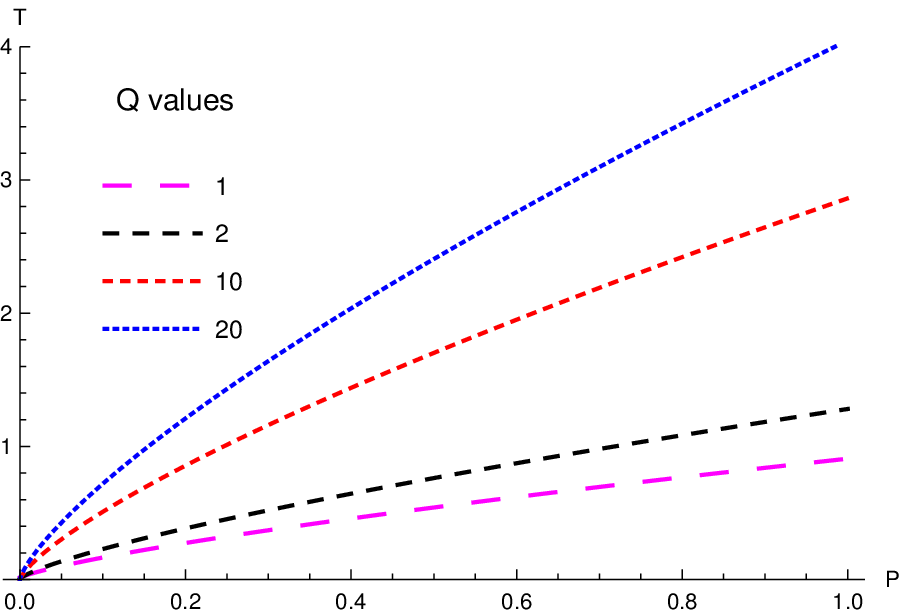}
    	\label{fig:4}
    }
    \\
    \subfigure[]
    {
    	\includegraphics[width=0.5\textwidth]{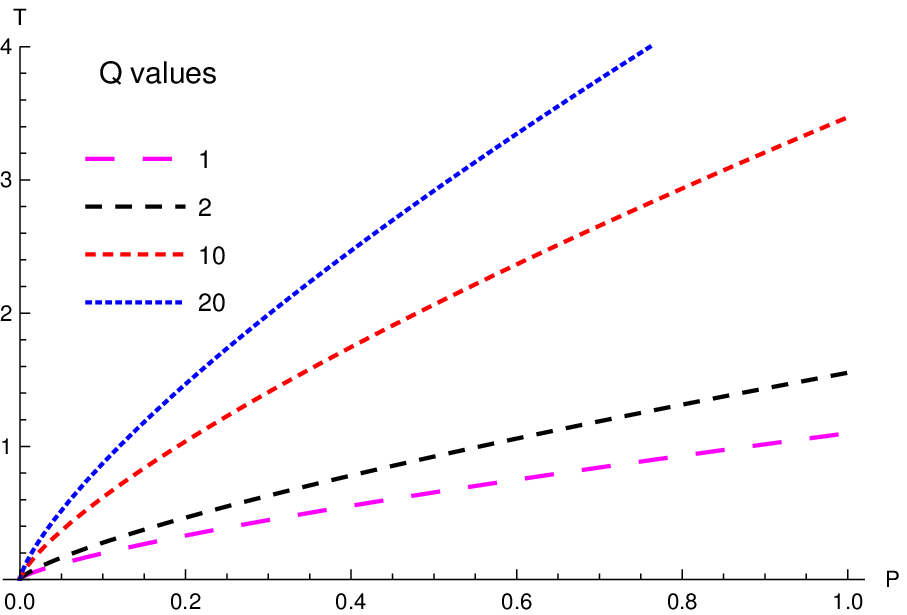}
    	\label{fig:5}
    }
    \subfigure[]
    {
    	\includegraphics[width=0.5\textwidth]{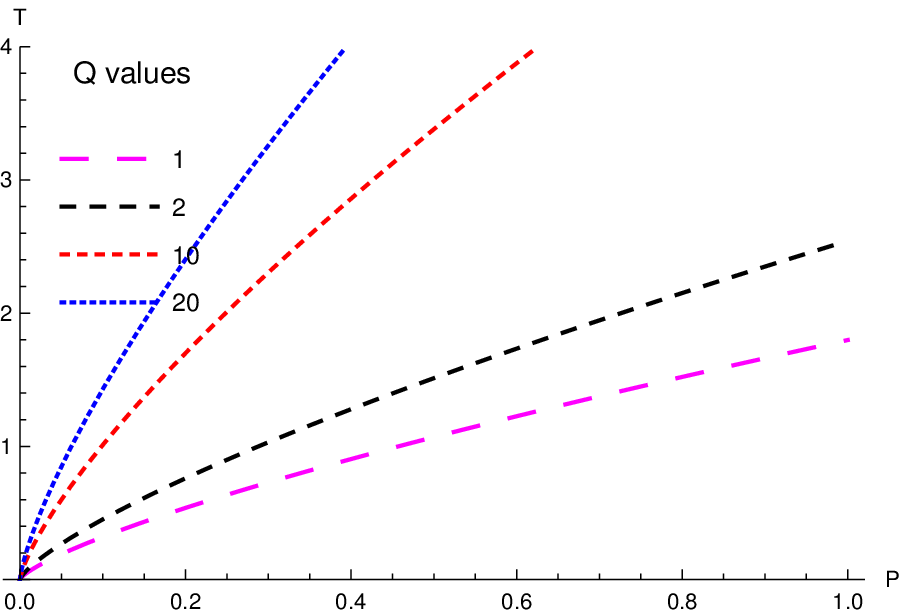}
    	\label{fig:6}
    }
  \caption{Inversion curves for charged AdS black hole with global monopole parameter $\eta =0, 0.1, 0.3, 0.5, 0.7, 0.9$ from top to bottom. The plots are the locus of inversion ponts $(P_i,T_i)$. Increasing $\eta$ increases the inversion temperature for fixed pressure.}
    \label{linear plots}
\end{figure*}
\begin{figure*}[!]
	\subfigure[]
	{
		\includegraphics[width=0.5\textwidth]{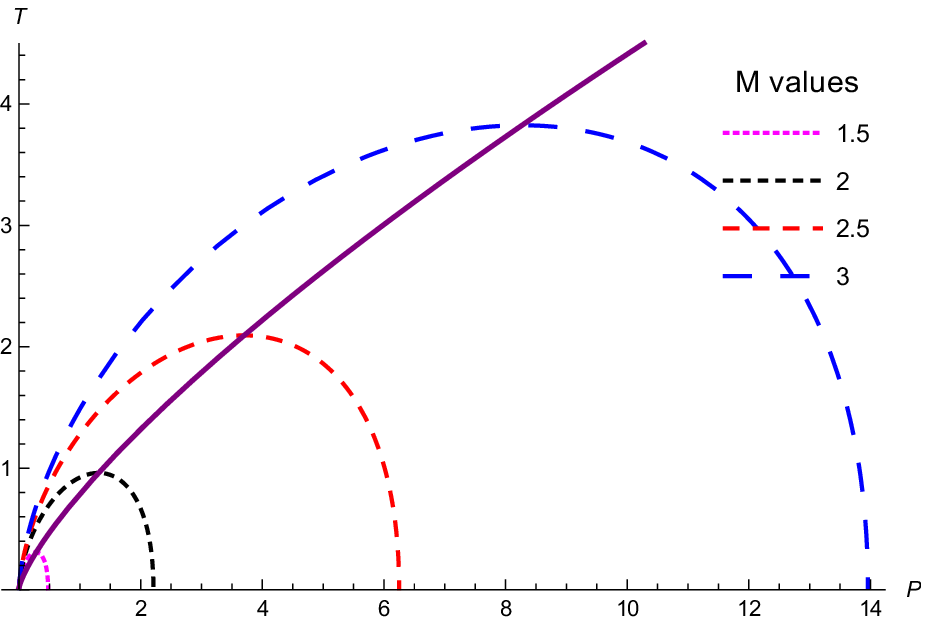}
		\label{fig:firstsub1}
	}
	\subfigure[]
	{
		\includegraphics[width=0.5\textwidth]{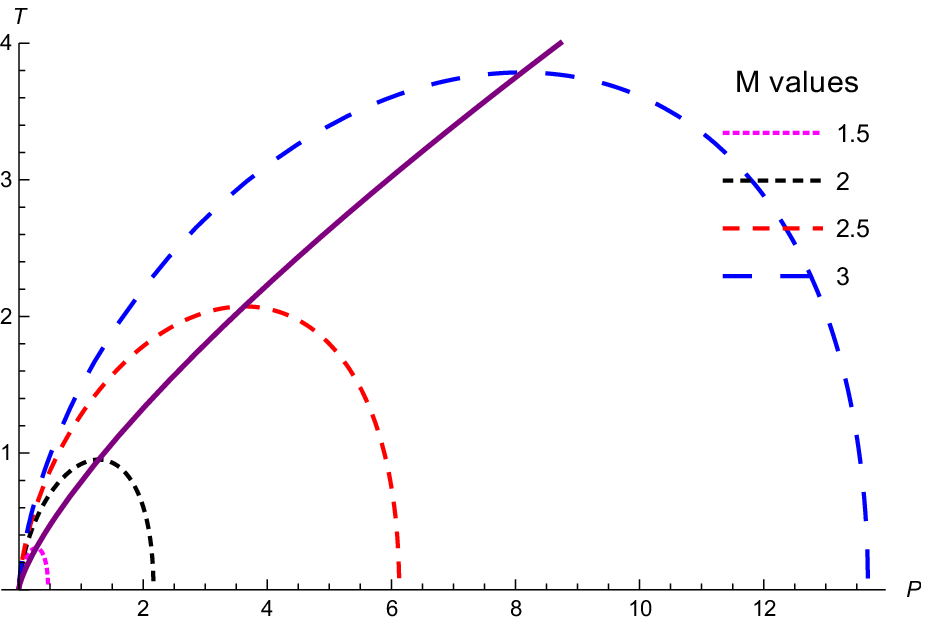}
		\label{fig:firstsub2}
	}
\\
    \subfigure[]
    {
        \includegraphics[width=0.5\textwidth]{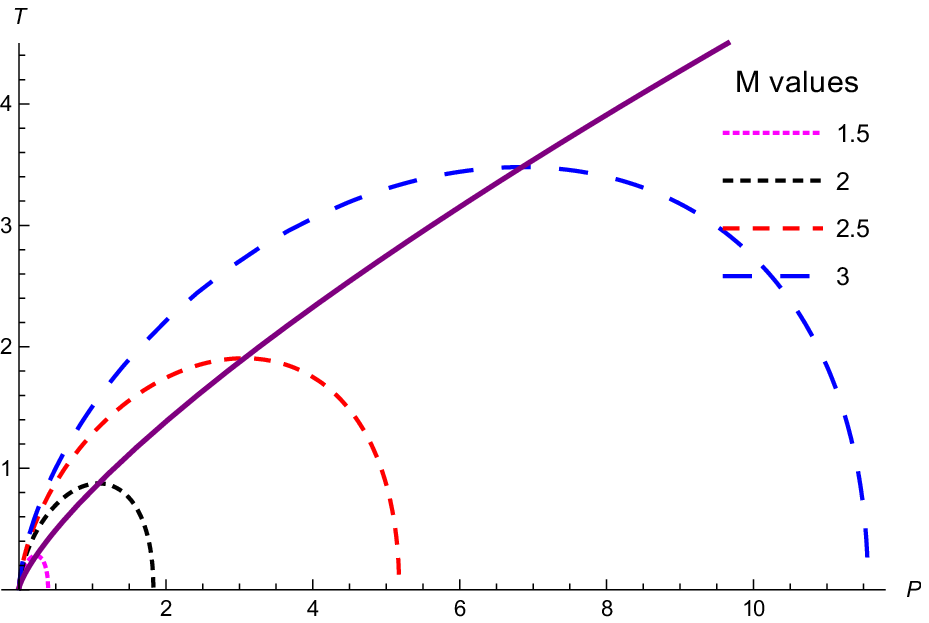}
        \label{fig:firstsub3}
     }
    \subfigure[]
    {
    	\includegraphics[width=0.5\textwidth]{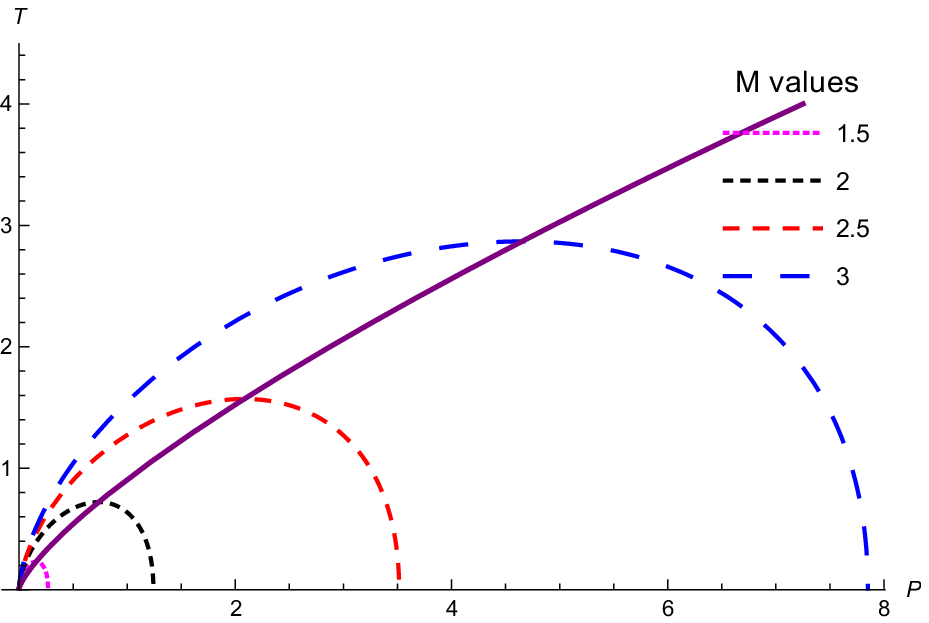}
    	\label{fig:secondsub4}
    }
\\
    \subfigure[]
    {
    	\includegraphics[width=0.5\textwidth]{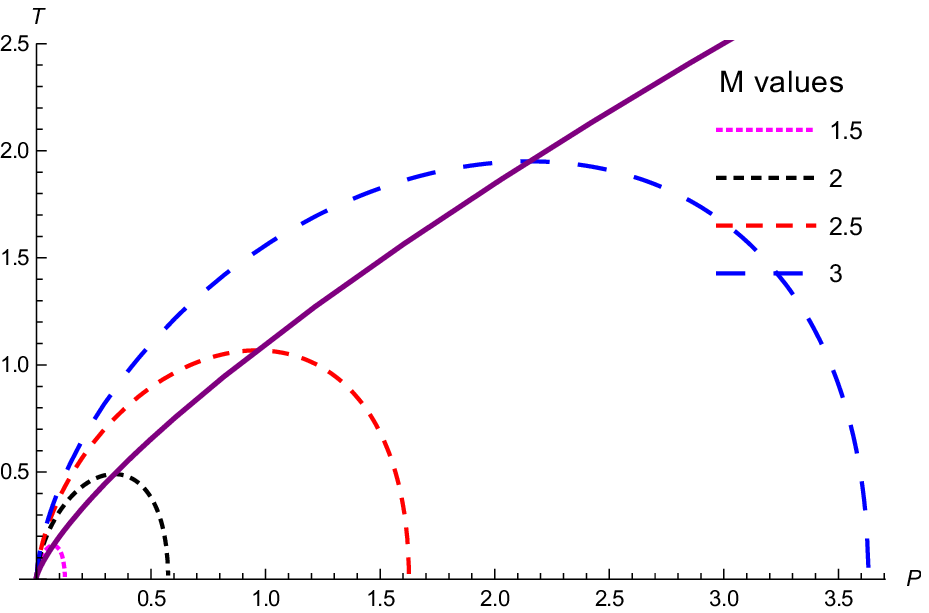}
    	\label{fig:secondsub5}
    }
	\subfigure[]
	{
		\includegraphics[width=0.5\textwidth]{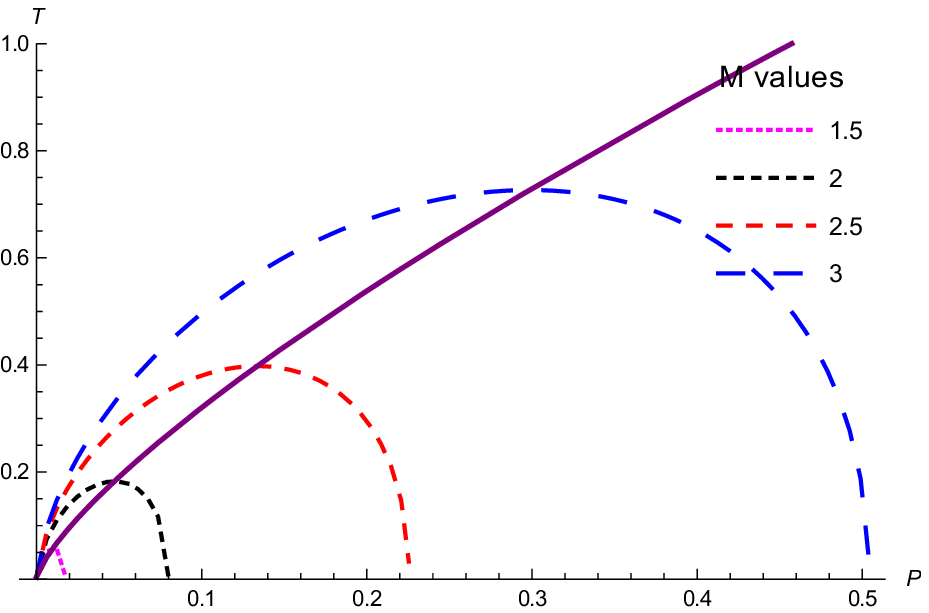}
		\label{fig:firstsub6}
	}
    \\
  \caption{Crossing diagrams between inversion and isenthalpic curves for different values of $\eta$. ($\eta = 0 , 0.1 , 0.3 , 0.5 , 0.7, 0.9$ from top to bottom and $Q=1$). }
    \label{crossing}
\end{figure*}
\newpage
\section{Conclusions}
In this paper, we have explored the Joule Thomson expansion for AdS black hole with global monopole in extended phase space. Firstly, we studied the thermodynamic properties of the black hole and analysed the effect of $\eta$ on these properties. Then the JT coefficient $\mu$ which determines the cooling and heating phase was obtained for van der Waals gas. Followed by the study of inversion and isenthalphic curves for the gas. 

We applied the idea of JT expansion to charged AdS black hole with global monopole where the key ingredient was the symmetry breaking parameter $\eta$. The tradtional JT coefficient analysis, isenthalpic and inversion curve studies were done for this metric with different values of $\eta$. The result is interesting since all the critical quantities $P_c$, $T_c$ and $V_c$ and equation of state (\ref{eqstate}) depends on $\eta$\cite{Deng2018}. The inversion temperature and pressure both increases monotonically with $\eta$, which is evident from the inversion and isenthalpic  curves. From the inversion and isenthalpic curves we conclude that the sensitivity of $T_i$ for a given value of $P_i$ with increasing $\eta$ is stringent. Presence of global monopole leads to larger increase in inversion temperature than inversion pressure, thus the Joule-Thomson coefficient $\mu$ has a drastic increase.

It is a well known fact that in early universe  the global monopole plays an interesting role in density fluctuations which leads to formation of galaxies  and clusters in several theoretical approaches \cite{Bennett90,Bennett93,Pando98}.  Hence we hope that the study of JT expansion with monopole term will be significant from cosmological perspective. As an extension of this work in future, we seek the significance of cooling and heating phase in JT expansion  related to cosmological transitions.

\section*{acknowledgments}

Author N.K.A. would like to thank U.G.C. Govt. of India for financial assistance
under UGC-NET-JRF scheme.

\bibliographystyle{unsrt}
\bibliography{BibTex}
\nocite{}

\end{document}